\documentclass[12pt]{iopart}

\usepackage{amsmath}
\usepackage{iopams}
\usepackage{graphicx}
\usepackage[square,sort&compress,comma,numbers]{natbib}

\begin{document}

\title[High-sensitivity extreme-ultraviolet transient absorption spectroscopy enabled by  ...]{High-sensitivity extreme-ultraviolet transient absorption spectroscopy enabled by machine learning}
\author{Tobias Heinrich}
\address{Max Planck Institute for Multidisciplinary Sciences, 37077 Göttingen, Germany}
\author{Hung-Tzu Chang}
\address{Max Planck Institute for Multidisciplinary Sciences, 37077 Göttingen, Germany}
\author{Sergey Zayko}
\address{Max Planck Institute for Multidisciplinary Sciences, 37077 Göttingen, Germany}
\address{4th Physical Institute – Solids and Nanostructures, University of Göttingen, 37077 Göttingen, Germany}
\author{Murat Sivis}
\address{Max Planck Institute for Multidisciplinary Sciences, 37077 Göttingen, Germany}
\address{4th Physical Institute – Solids and Nanostructures, University of Göttingen, 37077 Göttingen, Germany}
\author{Claus Ropers}
\address{Max Planck Institute for Multidisciplinary Sciences, 37077 Göttingen, Germany}
\address{4th Physical Institute – Solids and Nanostructures, University of Göttingen, 37077 Göttingen, Germany}
\ead{claus.ropers@mpinat.mpg.de}
\vspace{10pt}
\begin{indented}
\item[]April 2023
\end{indented}

\begin{abstract}
	We introduce a machine-learning-based approach to enhance the sensitivity of optical-extreme ultraviolet (XUV) transient absorption spectroscopy. A reference spectrum is used as input to a three-layer feed-forward neural network, enabling an efficient elimination of source noise from measurement data. In pump-probe experiments using high-harmonic radiation, we show a more than tenfold improvement in noise suppression in XUV transient absorption spectra compared to conventional referencing. Utilizing strong spectral correlations in the source fluctuations, the network facilitates a pixel-wise noise reduction without the need for wavelength calibration of the reference spectrum. The presented method can be adapted to a wide range of beam lines and enables the investigation of subtle electron and lattice dynamics in the weak excitation regime, relevant for the study of photovoltaics and photoinduced phase transitions of strongly correlated materials.
\end{abstract}

\maketitle
 
\newpage


Table-top optical-extreme-ultraviolet (optical-XUV) transient absorption spectroscopy is a powerful tool to investigate photoinduced electronic and structural dynamics in atoms, molecules, and solids \cite{krauszAttosecondPhysics2009,geneauxTransientAbsorptionSpectroscopy2019,liuElementspecificElectronicStructural2021,biswasExtremeUltravioletReflection2022}. In this approach, a sample material is excited by an optical pulse and subsequently probed in reflection or transmission with a time-delayed XUV pulse produced by high-harmonic generation (HHG). The XUV pulse excites core electrons, and thus enables element-specific probing of the optically induced dynamics with intrinsic attosecond timing stability due to the phase-locked generation \cite{huppertAttosecondBeamlineActively2015}. In addition, the HHG process allows for single isolated attosecond pulses enabling unprecedented time resolution for the observation of sub-cycle electron dynamics \cite{krauszAttosecondPhysics2009,krauszAttosecondMetrologyElectron2014,Zhao2012}.

Despite its wide range of applications, table-top optical-XUV transient absorption spectroscopy has been mainly utilized in the high-pump-fluence regime ($>$1 mJ/cm$^2$). This is largely due to the strong nonlinearities in HHG that greatly amplify fluctuations of the driving laser pulse, leading to a typical noise floor of 0.1 - 1 mOD in state-of-the-art experiments \cite{Geneaux2021,zurch2017direct,stoossXUVbeamlineAttosecondTransient2019,weisshauptUltrafastModulationElectronic2017,katoHighlySensitiveTransient2020}, which presents a challenge for the observation of small pump-induced XUV absorbance changes. This far-from-shot-noise-limited sensitivity hinders studies of dynamics in the low-excitation regime, as involved in carrier dynamics in photovoltaics and electronic and structural phase transitions in solids \cite{basovElectrodynamicsCorrelatedElectron2011,imadaMetalinsulatorTransitions1998,mohr-vorobevaNonthermalMeltingCharge2011}. 
For example, in a solar cell, the excited carrier density by sunlight is typically on the order of $10^{14}$-$10^{17}$ cm$^{-3}$ \cite{wuComparisonChargeCarrier2022,upretyPhotogeneratedCarrierTransport2019}, whereas the carrier density in a 200 nm thick silicon membrane excited by 800 nm laser irradiation with fluence of 1 mJ/cm$^{2}$ exceeds $10^{18}$ cm$^{-3}$ \cite{Cushing2019}. 

\begin{figure*}[htb!]
\centering
\includegraphics[width=\textwidth]{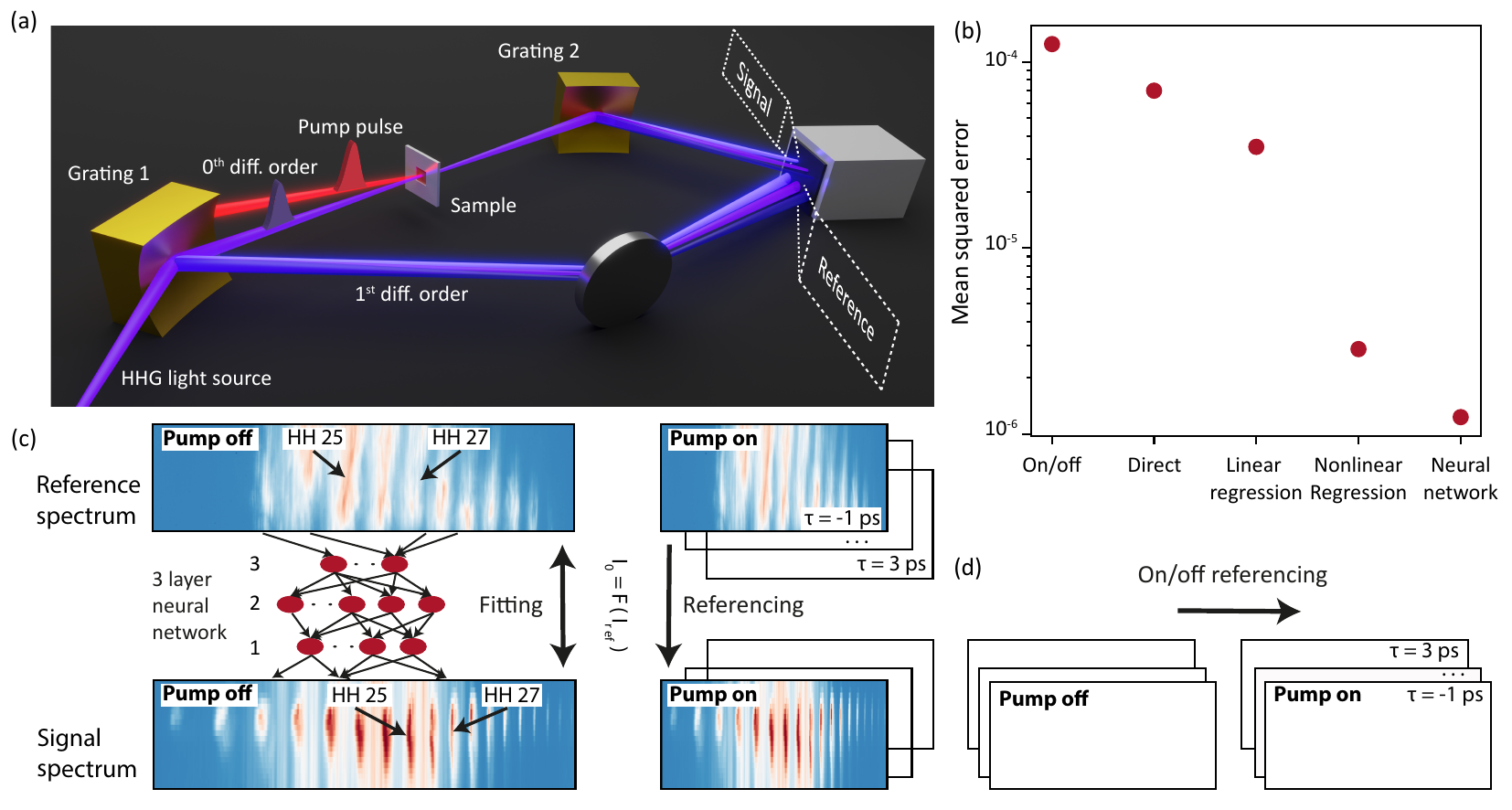}
\caption{(a) Experimental setup for extreme ultraviolet transient absorption spectroscopy with a reference spectrum. (b) Comparison of different referencing algorithms. (c) Scheme for referencing with a neural network, trained on the un-pumped data. The model is applied at every pump-probe time delay $\tau$. (d) Noise control scheme by comparing pumped and un-pumped signal (on/off) spectrum without additional reference spectrum.}
\label{Fig:1}
\end{figure*}
Substantial efforts have already been made to enhance the signal-to-noise ratio of transient absorption spectroscopy in the context of HHG sources. The most common method to trace long-term fluctuations involves a successive recording of pumped and un-pumped spectra \cite{stoossXUVbeamlineAttosecondTransient2019,zurch2017direct}. Additionally, noise reduction is possible by referencing, using the laser power \cite{volkovReductionLaserintensitycorrelatedNoise2019}, the spectral regions unaffected by the pump-induced dynamics \cite{Geneaux2021,faccialaRemovalCorrelatedBackground2021}, or a second spectrum of the XUV pulse before the sample \cite{willemsOpticalIntersiteSpin2020}. Although approaches with a second spectrum may largely overcome noise due to both long and short term fluctuations, imperfectly matched spectra can lead to far-from-ideal noise suppression.

In this work, we utilize machine learning to find the optimal referencing given an imperfect reference spectrum. Linear and nonlinear regression and a neural network model are evaluated for signal enhancement of experimental XUV transient absorption spectra. In experiments with a 60 nm thick 1T-TiSe$_2$ sample at low fluences (0.66 mJ/cm$^2$), we find that a three-layer neural network surpasses all other methods and enables studies of coherent lattice vibrations and a phase transition in this material \cite{HeinrichElectronicAndStructural}. Moreover, we exploit the capability of performing a pixel-wise referencing with sub harmonic spectral resolution using the neural network.

The experimental setup is illustrated in Fig.~\ref{Fig:1}(a). We investigate a 1T-TiSe$_2$ sample with XUV pulses produced by HHG with a 2 mJ, 35-fs-duration, 800-nm-wavelength driving pulse and its second harmonic from a Ti:sapphire laser operating at 1 kHz repetition rate. A second femtosecond optical pump pulse at 2 $\mu$m wavelength produced by optical parametric amplification from the same laser system is used to excite the sample. We employ discrete harmonics that are even and odd multiples of the 1.55 eV fundamental photon energy. A grating is placed before the sample, and the first-order diffracted XUV beam from the grating is taken as the \textit{reference} spectrum. The zeroth-order beam probes the sample and is dispersed by a second grating producing the \textit{signal} spectrum. The signal and the reference XUV beam illuminate the upper and lower half of a charged-coupled device (CCD, 1024 $\times$ 255 pixels), respectively. By recording the reference beam intensity $I_{ref}$, the noise of each acquisition of the signal spectrum $I_{sig}$ may be directly compensated. Transient absorption experiments measure the change of absorbance $\Delta A=-\log_{10} (I_{sig}/I_{0})$, defined as the logarithm of the pumped signal spectrum normalized by $I_{0}$ which corresponds to the transmitted probe spectrum without the pump. The time-dependent information of $I_{sig}$ is obtained by varying the delay $\tau$ between pump and probe pulse.

Most experiments use successive recording of pumped and unpumped (on/off) signal spectra ($I_{0}$ corresponds to a subsequently acquired signal spectrum without the pump) to trace fluctuations on timescales exceeding the camera  acquisition time (see Fig.~\ref{Fig:1}(d)). However, noise originating from fluctuations on the acquisition timescale can only be suppressed if $I_0$ and $I_{sig}$ are known simultaneously. In a two-spectrometer configuration, the normalization spectrum $I_0$ can be calculated by the simultaneously acquired reference spectrum $I_{0}=F(I_{ref})$. Here, $F$ is a function that relates the reference spectrum with the unpumped sample absorption. 

In its simplest form, $F$ merely contains the averaged ratio between the intensity of the unpumped signal and reference spectrum (direct referencing): 
\begin{equation} \label{eq:direct}
F(\lambda,I_{ref})=I_{ref}(\lambda)\times \langle I_{sig}(\lambda)/I_{ref}(\lambda) \rangle,
\end{equation}
with $\lambda$ denoting the wavelength. 
In practice, however, the signal and reference beams may be subjected to different noise levels which prevents successful signal extraction. For example, the reference and signal beams may vary drastically in intensity and spectral purity induced by the differences in optics and path lengths between the two beams. Furthermore, the two spectra may cover different spectral regions due to their respective beam paths and it can be very challenging to precisely calibrate CCD pixels and corresponding wavelength for both spectra at the same time. To address these issues, machine learning, which is capable of \textit{learning} intensity relations between different wavelengths \cite{Geneaux2021,faccialaRemovalCorrelatedBackground2021}, is utilized to establish the pixel-wise relationship between the signal and reference spectrum. The model function $F$ is trained by 58240 sets (80\%) of data to learn the relation between $I_{0}$ and $I_{ref}$. The training data is acquired without optical pump. The efficacy of the models are tested by another 14560 sets (20\%) of unpumped data. In the following, we introduce three different machine learning models: linear and nonlinear regression, and a three-layer neural network. The performance of the three approaches are evaluated by comparing their mean squared error to \textit{direct} referencing and on/off referencing using the same dataset (Fig.~\ref{Fig:1}(b)). Due to the discrete nature of the harmonics in our experiment and the described problems for \textit{direct} and on/off approaches, we use the individual harmonics (7 in the reference and 15 in the signal spectrum) as pixels in the comparison.

In the nonlinear regression model of degree $N$, the intensity of $I_{0}$ at pixel $s$ is expressed as
\begin{equation} 
    I_{0,s}=F(s,I_{ref})=a_0^s+\sum_{k=1}^N \sum_{r_1,\ldots,r_k} a_{r_1,\ldots,r_k}^s I_{ref,r_1}\times \cdots \times I_{ref, r_k}\, ,	
\end{equation}
where indices $r_i$ run through all pixels containing the reference beam and $a_0^s$ and $a_{r_0,\ldots,r_k}^s$'s are fitting coefficients. 
\begin{figure*}
\centering
\includegraphics[width=\textwidth]{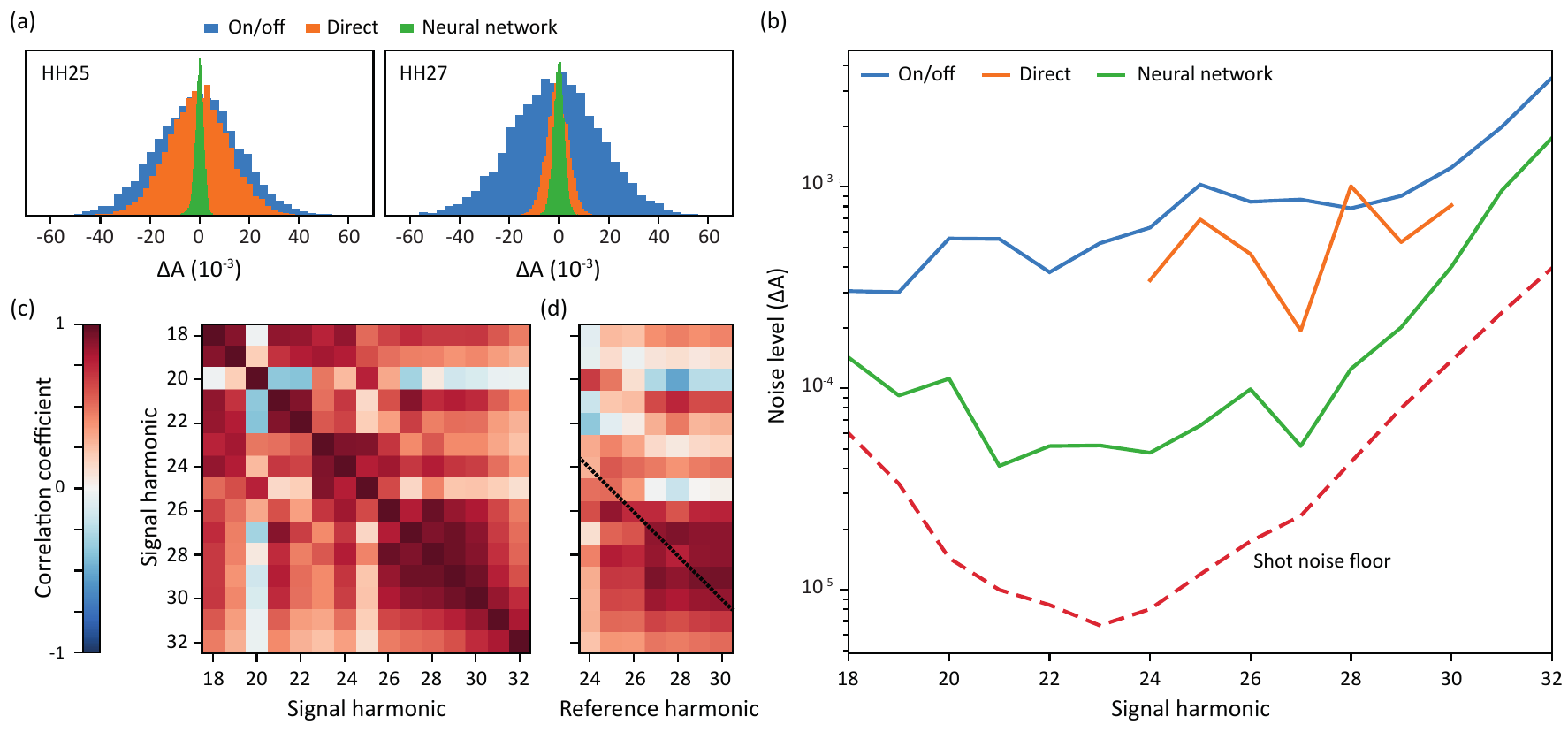}
\caption{(a) Histograms of optical density difference before time zero for harmonics 25 and 27. (b) Noise level for different referencing techniques derived by the histogram standard deviation as a function of the harmonic order. (c) Autocorrelation of the signal spectrum. (d) Cross-correlation between reference and signal spectrum. Diagonal elements (black line) are used for \textit{direct} referencing.}
\label{Fig:2}
\end{figure*}
The linear regression model uses $N=1$ with $(7+1) \times 15 = 120$ fitting parameters. As the number of fitting parameters increases exponentially with $N$, nonlinear regression models with different $N$'s are tested by 5-fold cross validation to avoid over-fitting \cite{cawleyOverfittingModelSelection2010}. Here, 20\% of the randomly ordered training data are permutatively taken out  to evaluate the degree of over-fitting. In this data set, over-fitting sets in at polynomial orders larger than 5 which correspond to 30030 fitting parameters.

The neural network model can be viewed as a further generalization, capable of fitting an arbitrary, non-polynomial function $F(I_{ref})$ to the data \cite{hornikMultilayerFeedforwardNetworks1989}. In the present work, we employ a three-layer feed-forward neural network (see Fig.~\ref{Fig:1}(c)) constructed with libraries \textit{Keras} and \textit{TensorFlow} \cite{chollet2015keras}. The code is available at \cite{HeinrichReferencingNeuralNetwork}. In the model, the number of neurons from hidden layers 1-3 are 250, 100, 70, and 200, 100, 20 for correlating intensity data of each pixel and each harmonic, respectively. These numbers have been empirically tested on the presented data and equivalent data sets recorded with the same setup and slightly changed HHG parameters. Larger models were found to yield no further accuracy improvements. A rectified-linear-unit activation function \cite{nairRectifiedLinearUnits2010,pmlr-v15-glorot11a} was used. The networks of 24435 and 193170 trainable parameters, corresponding to individual harmonic and pixel wise models, are optimized with stochastic gradient descent on the mean-squared error with the Adam optimizer \cite{kingmaAdamMethodStochastic2017} and batches of 50 data points \cite{keskarLargeBatchTrainingDeep2017}. Training of the larger model by 58240 data sets is achieved in less than three hours on a personal computer without parallelization, which presents a very moderate computational cost. In order to avoid over-optimization, the fitting is stopped when the error minimization of the testing data saturates. The presented architecture of a simple 3 layer neural network poses a good foundation for a model to describe the connection of signal and reference spectrum without any prior knowledge. In the future, more sophisticated designs may be employed to further improve the already great performance and accuracy.

Among the different models, the mean squared error (Fig.~\ref{Fig:1}(b)) of the pumped spectrum before time zero and a subsequently acquired unpumped spectrum (on/off) is the largest. Marginal improvements are obtained by \textit{direct} referencing (Eq.~(\ref{eq:direct})) and linear regression. The nonlinear regression method can reach better approximation with much smaller error, but is still surpassed by the neural network model. This indicates the importance to account for non-polynomial intensity noise imprinted by the highly nonlinear HHG process. To further analyze the performance of the different approaches, the models are applied to experimental XUV transient absorption data before time zero. Here, the pump pulse arrives \textit{after} the XUV probe and the transient absorption should therefore be zero. In Fig.~\ref{Fig:2}(a) the absorbance change of the 25$^\text{th}$ and 27$^\text{th}$ harmonic are compared between three de-noising procedures. Linear and nonlinear regression are not considered in this discussion as they are surpassed by the neural network within the machine learning context. The on/off approach shows the largest noise which is given by the standard deviation of the histograms. \textit{Direct} referencing improves the noise level, but is highly dependent on the harmonic order. This may result from various effects including clipping of some intensity, unequal focusing, and utilization of different optical elements with wavelength dependent reflection. Significantly better results are achieved with the neural network yielding similar noise levels at both harmonics. 

A comprehensive analysis of the noise level derived from the standard deviations as a function of harmonic order is presented in Fig.~\ref{Fig:2}(b). While the direct referencing improves the on/off approach at almost all wavelengths, the resulting noise strongly varies between harmonics. Please note that due to geometric constraints, only harmonic 24 to 30 can be recorded in the reference spectrum, limiting the spectral applicability of the direct referencing. The neural network shows significantly better performance with less deviation over the whole spectral region. We compare the results with the photon shot noise that defines the lower noise limit. At the utilized 200 msec integration time and 3 MHz readout rate, the dark current of 0.02 count and the readout noise of 30 counts are negligible in comparison to the measured signal of $>10^{5}$ counts per pixel. The referencing with a neural network reduces the noise level significantly, approaches the photon shot noise, and roughly matches its spectral dependence. The remaining discrepancy is likely to result form imperfect matching of signal and reference spectrum due to the different optical elements and beam paths. Furthermore, the spectral region which is not covered by the reference spectrum is prone to additional error.
\begin{figure*}
\centering
\includegraphics[width=\textwidth]{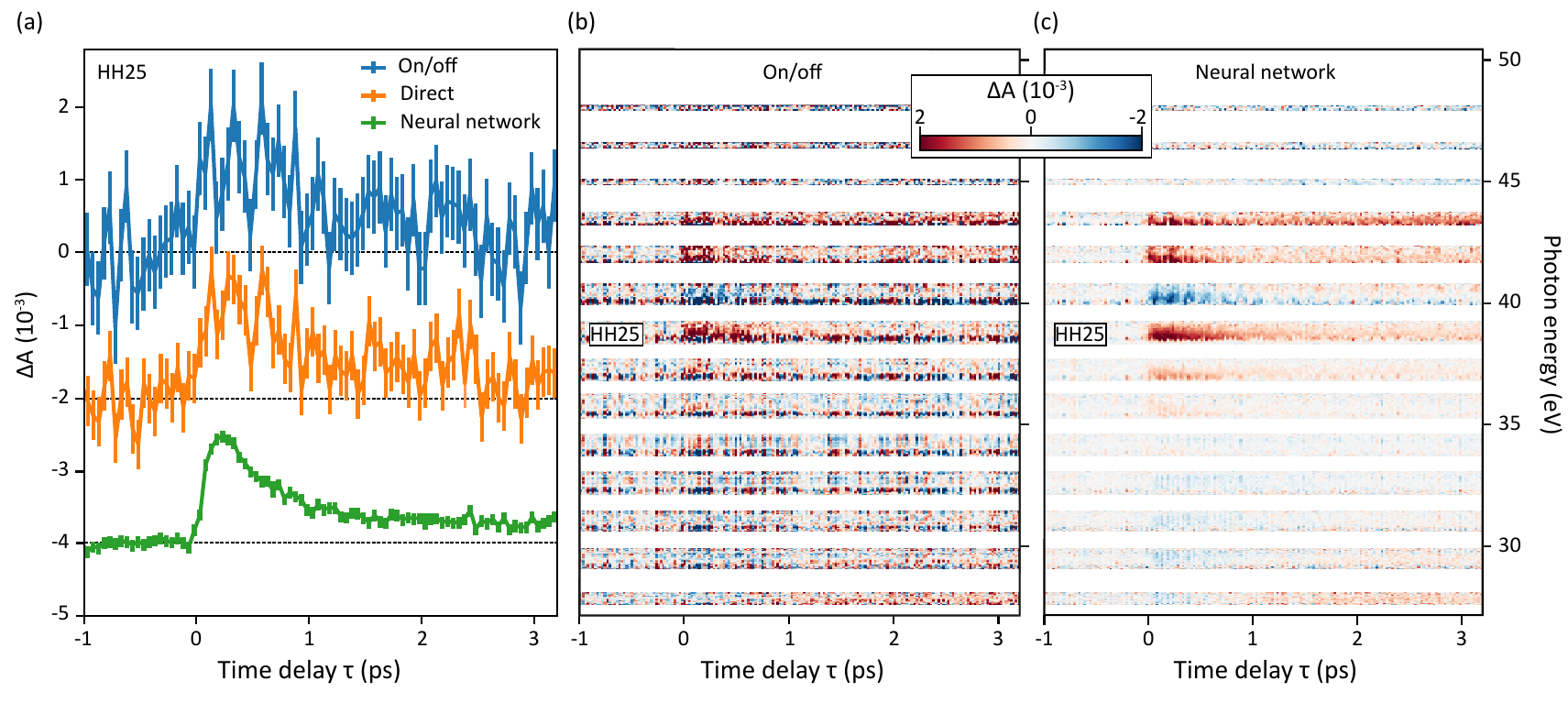}
\caption{(a) Transient absorption of the 25$^{\text{th}}$ harmonic evaluated with different referencing techniques. The error bars are identical for every delay and defined by the standard deviation at the particular harmonic. (b) Pixel wise spectrum obtained with the conventional on/off approach. (c) Same as (b) but with the neural network referencing.}
\label{Fig:3}
\end{figure*}

To validate the predictive power when the reference and signal spectra do not share the exact same wavelength ranges, we study the intensity correlation of different harmonic orders between the unpumped signal and reference spectra. Figures \ref{Fig:2}(c) and~\ref{Fig:2}(d) show the correlation coefficients among harmonics of the signal spectrum and between signal and reference spectrum, respectively. Most off-diagonal elements in both correlation maps have $> 0.4$ magnitude, indicating that the intensities of different harmonics are strongly correlated by the HHG process. The black line in Fig.~\ref{Fig:2}(d) shows components that are used in \textit{direct} referencing. As the intensities between different regions of the harmonic spectrum are highly correlated, it is feasible to predict the signal spectrum using a reference spectrum with partially overlapped spectral region. In addition, the off diagonal components provides additional information, which are used by the linear regression model. Nonlinear correlations fitted by the nonlinear regression and the neural network model are not captured by the correlation coefficients. In principle, any other correlated quantity, like laser power, HHG source variables or total XUV intensity could be used for de-noising with machine learning.

Finally, we demonstrate the signal enhancement capability by showing a pump-probe trace of the 25$^{\text{th}}$ harmonic for on/off, \textit{direct} and neural network referencing in Fig.~\ref{Fig:3}(a). The $9.4 \times 10^{-4} \, \text{OD}$ standard deviation of the spectrum with on/off referencing aligns with the typical value of $\approx 1$ mOD sensitivity found in most transient absorption studies \cite{Geneaux2021} and can be improved by a factor of two with \textit{direct} referencing. An improvement by more than one order of magnitude in sensitivity to $9 \times 10^{-5}$ OD is achieved with the neural network. In addition, the neural network model can be applied to obtain a pixel-wise correlation between the signal and reference spectra. As most attosecond sources offer a broadband supercontinuum covering large portions of the XUV region, pixel-wise correlation offers noise-reduction without the loss of spectral resolution due to binning. Transient absorption spectra as a function of pump-probe delays and XUV photon energies are compared in Fig.~\ref{Fig:3}(b) and Fig.~\ref{Fig:3}(c) between the established on/off approach and the neural network referencing, respectively. For clarity, the spectral regions with low XUV intensities (owing to the discrete nature of harmonics) are omitted. While the pump induced response remains at a similar level, the neural network referencing drastically improves the visibility of the pump probe data. Again one order of magnitude improvement down to roughly $10^{-4}$ OD is visible. Note that while nonlinear regression can also handle nonlinearities in the spectral correlation, it is not able to provide such single pixel spectra due to the large number of fit parameters. For example, in the case of a 1024 pixel detector, $10^{14}$ fit parameters would be needed to model a polynomial of 5$^{\text{th}}$ degree in nonlinear regression.

The feasibility of single pixel referencing renders the neural network algorithm ideally suited for normalizing the broadband spectra from a wide range of attosecond beam lines. In addition, no wavelength calibration or complete coverage of the signal spectrum is needed. Furthermore, any measured quantity which is correlated to the harmonic intensity such as laser power, total XUV intensity, pulse duration, pulse shape, gas pressure or beam position may be use as input in the machine learning algorithm. As a side note, the neural network should always be trained during the data acquisition on the unpumped data such that the model can accurately capture fluctuations of the HHG source during the measurement. Comparing with the widely used on/off referencing, the neural network method does not require additional experiment time since pumped data are recorded by default for the on/off referencing in quick succession to un-pumped spectra. Note that when using a pre-trained model for noise-reduction of datasets in a different experiment, the performance of the model may degrade due to differences in HHG parameters.
 
In summary, we developed an algorithm for optimal noise correction in transient absorption spectroscopy with a reference spectrum. Thereby, we extend the applications of data driven algorithms in optics and photonics \cite{gentyMachineLearningApplications2021} which already showed great success in X-ray diffraction and spectroscopy for noise suppression and signal extraction \cite{chenMachineLearningNeutron2021}. It was found that a three-layer neural network approach improves the noise level by an order of magnitude compared to the established on/off referencing method. This sensitivity increase opens the field of femtosecond to attosecond XUV transient absorption spectroscopy in the weak excitation regime such as photoinduced phase transitions in photovoltaics \textit{in operando} or strongly correlated materials like 1T-TiSe$_2$. As the neural network approach utilizes the nonlinear intensity correlation between different pixels, it does not require wavelength-calibrated reference spectra, nor does it rely on signal and reference beams covering the same spectral region. This largely relaxes the conditions for experimental instrumentation and highlights the wide application of our method to both narrowband and broadband sources in the field of transient spectroscopy. On a broader scale, the presented combination of referencing with machine learning may be applied to a large variety of experiments with fluctuating probes to enhance the sensitivity.

 \ack 
This work was funded with resources from the Gottfried Wilhelm Leibniz Prize and the Deutsche Forschungsgesellschaft (DFG). H.-T. C. acknowledges support from the Alexander von Humboldt Foundation.


\bibliographystyle{iopart-num}
\bibliography{references}
\end{document}